\documentclass[useAMS,usenatbib]{mn2e}
\usepackage{multirow}
\usepackage{epsfig}
\usepackage{subfigure}

\makeatletter
\let\jnl@style=\rm
\def\ref@jnl#1{{\jnl@style#1}}

\def\aj{\ref@jnl{AJ}}                   
\def\araa{\ref@jnl{ARA\&A}}             
\def\apj{\ref@jnl{ApJ}}                 
\def\apjl{\ref@jnl{ApJ}}                
\def\apjs{\ref@jnl{ApJS}}               
\def\ao{\ref@jnl{Appl.~Opt.}}           
\def\apss{\ref@jnl{Ap\&SS}}             
\def\aap{\ref@jnl{A\&A}}                
\def\aapr{\ref@jnl{A\&A~Rev.}}          
\def\aaps{\ref@jnl{A\&AS}}              
\def\azh{\ref@jnl{AZh}}                 
\def\baas{\ref@jnl{BAAS}}               
\def\jrasc{\ref@jnl{JRASC}}             
\def\memras{\ref@jnl{MmRAS}}            
\def\mnras{\ref@jnl{MNRAS}}             
\def\pra{\ref@jnl{Phys.~Rev.~A}}        
\def\prb{\ref@jnl{Phys.~Rev.~B}}        
\def\prc{\ref@jnl{Phys.~Rev.~C}}        
\def\prd{\ref@jnl{Phys.~Rev.~D}}        
\def\pre{\ref@jnl{Phys.~Rev.~E}}        
\def\prl{\ref@jnl{Phys.~Rev.~Lett.}}    
\def\pasp{\ref@jnl{PASP}}               
\def\pasj{\ref@jnl{PASJ}}               
\def\qjras{\ref@jnl{QJRAS}}             
\def\skytel{\ref@jnl{S\&T}}             
\def\solphys{\ref@jnl{Sol.~Phys.}}      
\def\sovast{\ref@jnl{Soviet~Ast.}}      
\def\ssr{\ref@jnl{Space~Sci.~Rev.}}     
\def\zap{\ref@jnl{ZAp}}                 
\def\nat{\ref@jnl{Nature}}              
\def\iaucirc{\ref@jnl{IAU~Circ.}}       
\def\aplett{\ref@jnl{Astrophys.~Lett.}} 
\def\apspr{\ref@jnl{Astrophys.~Space~Phys.~Res.}}
\def\bain{\ref@jnl{Bull.~Astron.~Inst.~Netherlands}}
\def\fcp{\ref@jnl{Fund.~Cosmic~Phys.}}  
\def\gca{\ref@jnl{Geochim.~Cosmochim.~Acta}}   
\def\grl{\ref@jnl{Geophys.~Res.~Lett.}} 
\def\jcp{\ref@jnl{J.~Chem.~Phys.}}      
\def\jgr{\ref@jnl{J.~Geophys.~Res.}}    
\def\jqsrt{\ref@jnl{J.~Quant.~Spec.~Radiat.~Transf.}}
\def\memsai{\ref@jnl{Mem.~Soc.~Astron.~Italiana}}
\def\nphysa{\ref@jnl{Nucl.~Phys.~A}}   
\def\physrep{\ref@jnl{Phys.~Rep.}}   
\def\physscr{\ref@jnl{Phys.~Scr}}   
\def\planss{\ref@jnl{Planet.~Space~Sci.}}   
\def\procspie{\ref@jnl{Proc.~SPIE}}   

\makeatother

\title[NGC~3147: a ``true'' Seyfert 2 without the broad-line region]{NGC~3147: a ``true'' Seyfert 2 without the broad-line region}

\author[Stefano Bianchi, et al.]{S. Bianchi$^1$\thanks{E-mail: bianchi@fis.uniroma3.it (SB)}, A. Corral$^2$, F. Panessa$^2$, X. Barcons$^2$, G. Matt$^1$, L. Bassani$^3$,
\newauthor
F.J. Carrera$^2$, E. Jim\'enez-Bail\'on$^{4,5}$\\
$^1$Dipartimento di Fisica, Universit\`a degli Studi Roma Tre, via della Vasca Navale 84, 00146 Roma, Italy\\
$^2$Instituto de F\'\i sica de Cantabria (CSIC-UC), 39005 Santander, Spain\\
$^3$INAF-IASF, Via P. Gobetti 101, 40129 Bologna, Italy\\
$^4$Instituto de Astronom\'ia, Universidad Nacional Aut\'onoma de M\'exico, Apartado Postal 70-264, 04510 Mexico DF, Mexico\\
$^5$LAEFF Apd. 78 Villanueva de la Ca\~nada - 28691-Madrid, Spain
}

\begin{document}


\maketitle

\label{firstpage}

\begin{abstract}
We report on simultaneous optical and X-ray observations of the
Seyfert galaxy, NGC~3147. The XMM-\textit{Newton} spectrum
shows that the source is unabsorbed in the X-rays
($N_H<5\times10^{20}$ cm$^{-2}$). On the other hand, no broad lines
are present in the optical spectrum. The origin of this
optical/X-rays misclassification (with respect to the Unification Model) cannot be attributed to variability, since the observations in the two bands are simultaneous.
Moreover, a Compton-thick nature of the object can be rejected on
the basis of the low equivalent width of the iron K$\alpha$ line
($\simeq130$ eV) and the large ratio between the 2-10 keV and the
[O\textsc{iii}] fluxes. It seems therefore inescapable to conclude that NGC~3147 intrinsically lacks the Broad Line Region (BLR), making it the first ``true'' Seyfert 2. 

\end{abstract}

\begin{keywords}
galaxies: active - galaxies: Seyfert - X-rays: individual: NGC3147
\end{keywords}

\section{Introduction}

The basic assumption of Unified Models of Active Galactic Nuclei
(AGN) is that type 1 and type 2 objects are intrinsically the same,
the apparent difference being solely due to orientation effects
\citep[e.g.][]{antonucci93}. The absorbing medium assumes the fundamental role in this scenario. It is usually envisaged as an
optically thick `torus', embedding the nucleus and the Broad Line
Region (BLR). If we observe the torus edge-on, all the nuclear
radiation, as well as the broad optical lines coming from the BLR, is completely blocked and we classify the source
as a type 2. The narrow lines are still visible, because the Narrow
Line Region (NLR) is located farther away from the nucleus, beyond
the torus.  On the other hand, if the torus does not intercept our
line of sight, we observe every component of the spectrum and the
object is classified as a type 1.

Simple extrapolations of the optical/UV scenario to the X-ray
emission do not, however, always easily fit the observations,
leading sometimes to different classifications between the two
bands. A larger amount of absorbing material is usually measured
from X-ray observations in Seyfert 2s with respect to Seyfert 1s, as
expected \citep[e.g.][]{awaki91,ris99}.
However, a number of ``true'' Seyfert 2 candidates have been found, i.e. objects with no absorption in the X-rays and no broad optical lines \citep[e.g.][]{pappa01,pb02,wol05}, or with a large intrinsic Balmer decrement of the BLR \citep{bcc03,corr05}. The so-called `naked' AGN may be similar objects, being characterised by the absence of broad lines, but strong
variability in the optical band \citep{hawk04} and, apparently, no
absorption in the X-rays \citep{gliozzi07}. These findings represent
a challenge to the Unified Models and may require new classes
of objects with intrinsic differences, like the absence of
BLR.

Nevertheless, these sources may be highly variable and may change their optical and/or X-ray appearance
in different observations. These `changing-look' AGN are not
uncommon. In some cases, this behaviour is best explained by a real
`switching-off' of the nucleus \citep[see e.g.][]{mgm03,gua05}, in others
by a variable column density of the absorber
\citep[e.g.][]{elvis04,ris05}. If the optical and the X-ray spectrum
are taken in two different states of the source, it is clear that
the disagreement between the two classifications may be only
apparent. Therefore, the key to find genuine `unabsorbed Seyfert 2s'
is represented by simultaneous X-ray and optical observations.

NGC~3147 (z=0.00941) belongs to the Palomar optical spectroscopic
survey of nearby galaxies \citep{hfs95}. Its classification as a
Seyfert 2 is based on both the relative strength of the low-ionization
optical forbidden lines with respect to the hydrogen Balmer lines and the ratio of [O\textsc{iii}] to H$\beta$,
together with the lack of broad permitted lines \citep{hfs97}.
\textit{ASCA} provided the first X-ray spectrum, which appeared
Seyfert 1-like, without significant absorption and a standard
powerlaw index \citep{ptak96}. The lack of obscuration was later
confirmed by \textit{BeppoSAX} \citep{dad07} and \textit{Chandra},
which also showed that no off-nuclear source can significantly
contribute to the nuclear emission \citep{tw03}. In order to
reconcile the X-ray data with the optical classification,
\citet{ptak96} suggested that NGC~3147 was a Compton-thick source,
given also the relatively large equivalent width (EW) of the iron
line. However, this hypothesis was rejected by the use of diagnostic
diagrams based on $F_\mathrm{X}/F_{\mathrm{[OIII]}}$ and
$F_\mathrm{X}/F_{\mathrm{IR}}$ ratios \citep{pb02}. NGC~3147 is,
therefore, a genuine candidate to be an unabsorbed Seyfert 2 galaxy.
In this paper, we report on simultaneous XMM-\textit{Newton} and
optical observations of NCG~3147, in order to settle the issue.

In the following, errors correspond to the 90\% confidence level for
one interesting parameter ($\Delta \chi^2 =2.71$), where not
otherwise stated. The adopted cosmological parameters are $H_0=70$
km s$^{-1}$ Mpc$^{-1}$, $\Omega_\Lambda=0.73$ and $\Omega_m=0.27$ (i.e. the
default ones in \textsc {Xspec} 12.3.1).

\section{Data reduction}

\subsection{XMM-\textit{Newton}}

NGC~3147 was observed by XMM-\textit{Newton} on 2006, October 6th
(\textsc{obsid: 0405020601}). The observation was performed with the
EPIC CCD cameras, the pn and the two MOS, operated in Large Window and Thin Filter. Data
were reduced with \textsc{SAS} 7.0.0 and screening
for intervals of flaring particle background was done consistently
with the choice of extraction radii, in an iterative process based
on the procedure to maximize the signal-to-noise ratio described by
\citet{pico04}. After this process, the net exposure time was of
about 14, 16 and 17 ks for pn, MOS1 and MOS2 respectively, adopting
extraction radii of 40 arcsec for all the cameras. The background
spectra were extracted from source-free circular regions with a
radius of 50 arcsec. Pattern 0 to 4 were used for the pn spectrum,
while MOS spectra include patterns 0 to 12. Since the two MOS
cameras were operated with the same mode, we co-added MOS1 and MOS2
spectra, after having verified that they agree with each other and
with the summed spectrum. Spectra were binned in order to
oversample the instrumental resolution by at least a factor of 3 and
to have no less than 25 counts in each background-subtracted
spectral channel. The latter requirement allows us to use the
$\chi^2$ statistics. No significant variability is found in the lightcurve of the observation.

\subsection{Optical data}

NGC3147 was observed at the {\it Observatorio de Sierra Nevada} (OSN,
Granada, Spain) on 2006, October 4th, 5th and 9th. A total of 6 spectra
were taken with the spectrograph {\it Albireo}, with single exposure
times of 1800{\rm s}. The slit width was adjusted to a seeing of $\sim$ 2
arcsec and aligned with the galaxy's major axis. The spectral coverage
ranges from 4000 to 7000 \AA, with an average Full Width at Half Maximum (FWHM) spectral resolution
of $\sim4$\AA, measured from unblended arc lines. No significant spectral variations are present among
the spectra but, given the large air masses for the last two days (above 2), we only used the first night spectrum for our analysis (air
mass $\sim1.74$).

In order to extract the spectrum from the active nucleus with the
minimal contamination from the host galaxy, we constructed a spectral
template to mimic the latter. We extracted a central spectrum with an
aperture width of $\sim1$ arcsec. Then, we extracted the host galaxy
spectrum $\sim3$ arcsec away from the centre and the same aperture as
the central spectrum, in order to use it as one of our template
ingredients. \citet{moll04} showed that the disk of NGC~3147 has a
larger central surface brightness than the bulge, so the bulge is
a minor (but significant) component to the host galaxy at the
centre. To account for the bulge contribution at the centre, we therefore added to the host galaxy spectrum a
reddened elliptical template from the \citet{kin96} atlas. Then, we subtracted the resulting spectrum from the nuclear
spectrum so that the absorption bands (G-band, Mg I and Ca I) disappeared.

The subtracted spectrum was finally brought to an absolute flux by
scaling it in such a way that the [O\textsc{iii}]$\lambda$5007 line flux is
in agreement with the value measured by \citet{hfs97}. Standard reduction techniques were applied, including wavelength calibration against calibrated arc
spectrum and an approximate flux calibration using the
spectrophotometric standard star G191B2B. There was no attempt, however, to correct for aperture effect, so this calibration is only
good in relative terms.\\

\section{Data analysis}

\subsection{\label{xmmanalysis}XMM-\textit{Newton}}

\begin{figure*}
\epsfig{figure=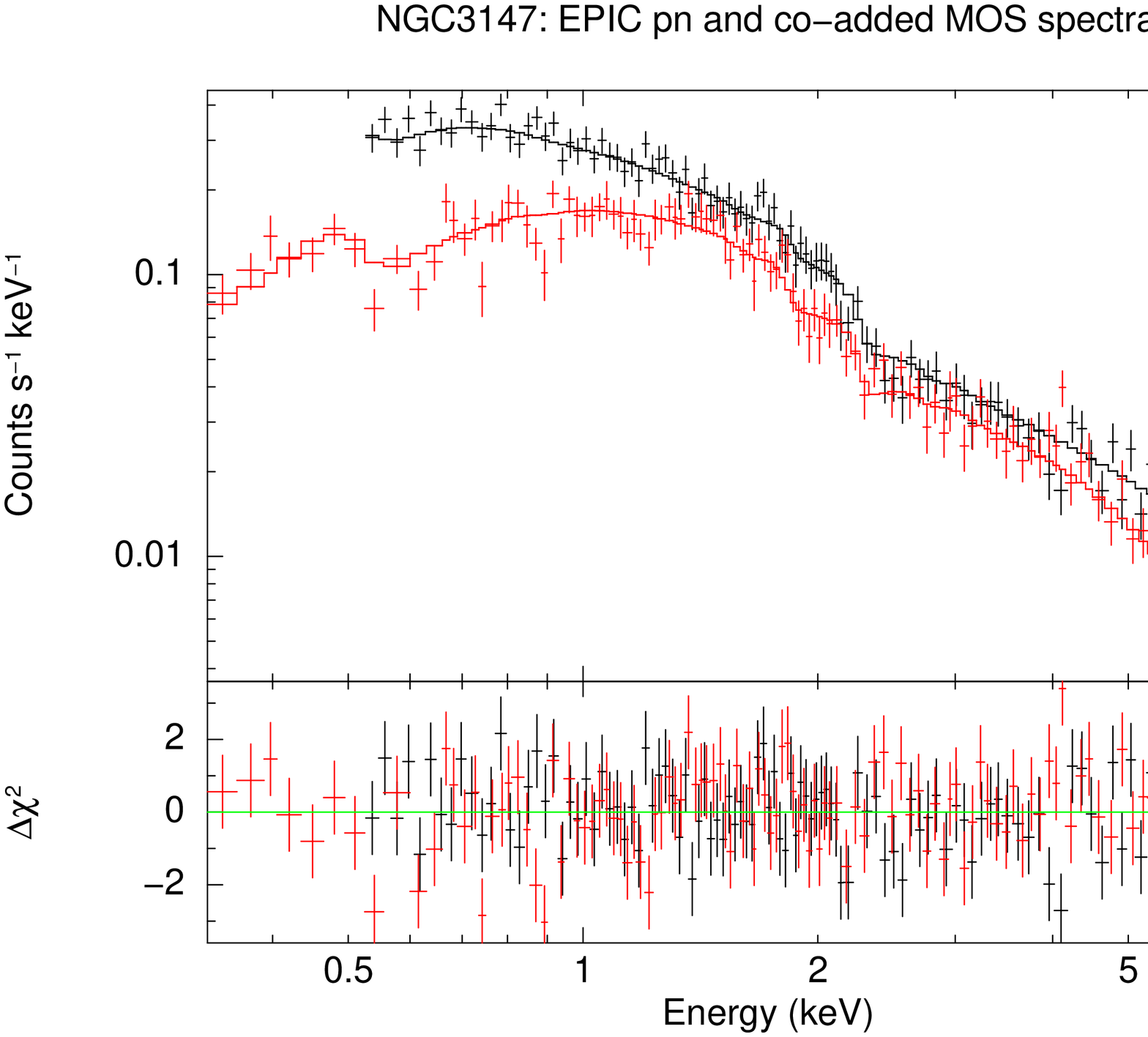,height=5.6cm}
\hspace{2cm}
\epsfig{figure=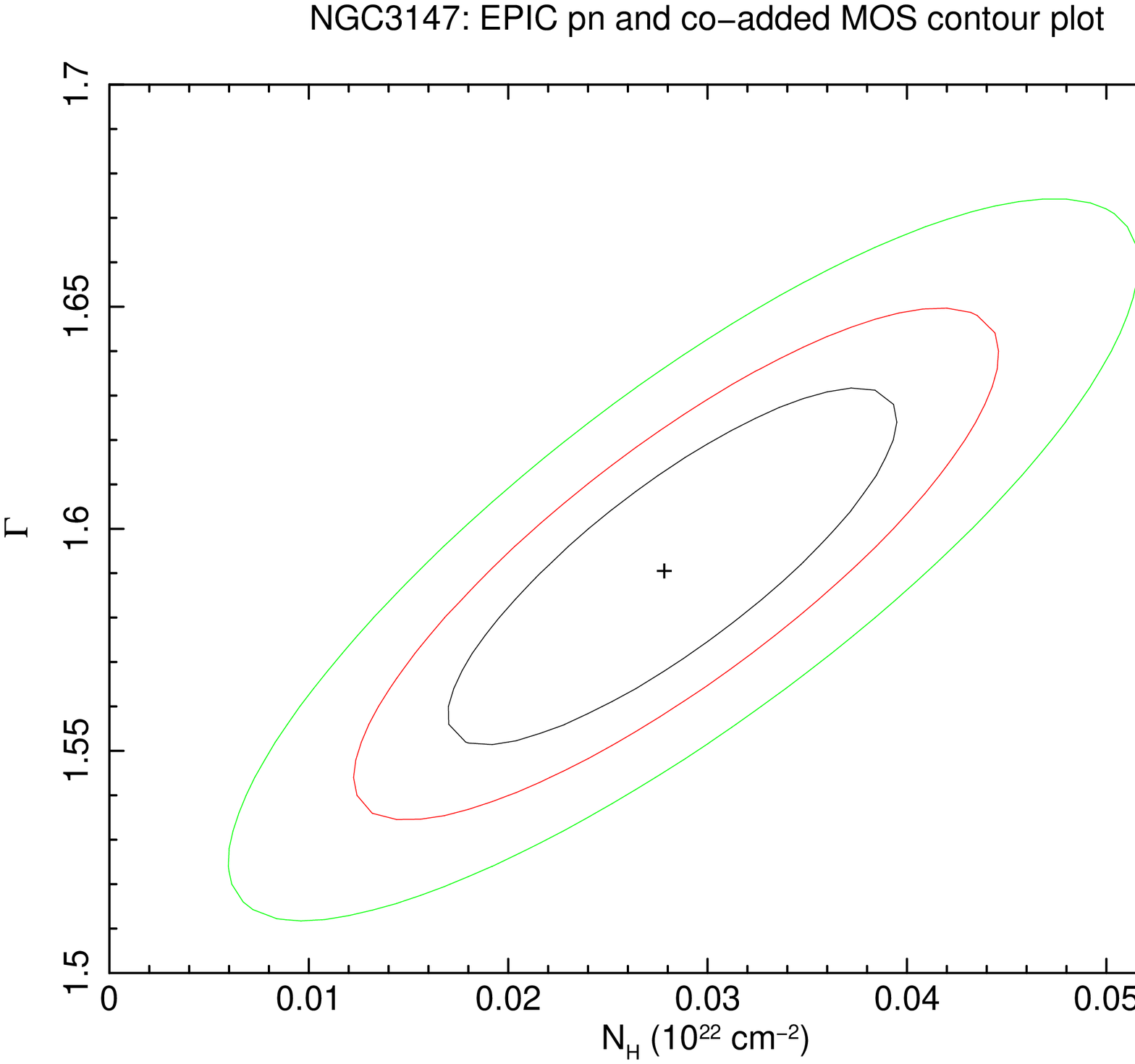,height=5.6cm}
\caption{\label{xmm}XMM-\textit{Newton} observation of NGC~3147: \textit{Left panel}: The EPIC pn and co-added MOS spectra, best fit models and $\Delta\chi^2$ deviations. \textit{Right panel}: Photon index vs. column density contour plots at 68, 90 and 99\% confidence level for two interesting parameters. The latter parameter does not include Galactic column density, which is fixed as a separate component. See text for details.}
\end{figure*}

The pn and co-added MOS spectra are well fitted by a simple powerlaw
($\Gamma\simeq1.6$), absorbed by the Galactic column density
\citep[$3.6\times10^{20}$ cm$^{-2}$: ][]{dl90}. Local absorption at
the redshift of the source is also required by the data, but the
column density is very low, being only
$\left(2.8\pm1.2\right)\times10^{20}$ cm$^{-2}$. As shown in the
right panel of Fig. \ref{xmm}, it is lower than $5\times10^{20}$
cm$^{-2}$ at the 99\% confidence level for two interesting
parameters. An emission line is clearly present, with a resolved
line width of $\sigma=0.27^{+0.28}_{-0.12}$ keV, a centroid energy
of $6.49^{+0.19}_{-0.15}$ keV and EW=$310^{+190}_{-70}$ keV.
However, the centroid energy and the large $\sigma$ suggests that
this is likely a blend of a neutral iron line and emission from
ionised iron, as often found in other Seyfert galaxies
\citep[e.g.][]{bianchi05}. Indeed, two emission lines, from neutral
iron and Fe \textsc{xxv}, are required by the data, at significance
of 99.3\% and 96\%, respectively, according to F-test\footnote{In principle, the
F-test is not a reliable test for the significance of emission
lines, but it can be used if their normalizations are allowed to be negative \citep{prot02}.}, with EW of around 130 eV for both of
them. An upper limit of 70 eV can be derived for a Fe \textsc{xxvi}
line. The final fit is as good\footnote{We note that significant
residuals are visible around 0.9 keV in the MOS co-added spectrum,
as well as the single MOS1 and MOS2 spectra, affecting the quality
of the fit. However, they are not present in the higher
signal-to-noise-ratio pn spectrum.} as the one with a broad line
(reduced $\chi^2$=1.1 for 215 d.o.f.): see Fig. \ref{xmm} and Table
\ref{xmmfit}. 

\begin{table}
\caption{\label{xmmfit}Best fit parameters for the XMM-\textit{Newton} observation of NGC~3147. Emission lines' energies were kept fixed.}
\begin{center}
\begin{tabular}{ll}
\hline
$N_{\mathrm{H}}$ (cm$^{-2}$)& $\left(2.8\pm1.2\right)\times10^{20}$\\
$\Gamma$ & $1.59\pm0.04$\\
$I_{6.4}$ (ph cm$^{-2}$ s$^{-1}$) & $\left(2.1\pm1.3\right)\times10^{-6}$\\
EW$_{6.4}$ (eV) & $130\pm80$\\
$I_{6.7}$ (ph cm$^{-2}$ s$^{-1}$) & $1.7^{+1.3}_{-1.5}\times10^{-6}$\\
EW$_{6.7}$ (eV) & $130^{+80}_{-100}$\\
$I_{6.96}$ (ph cm$^{-2}$ s$^{-1}$) & $<1.1\times10^{-6}$\\
EW$_{6.96}$ (eV) & $<70$\\
$F_\mathrm{0.5-2\,keV}$ (cgs) & $\left(5.7\pm0.3\right)\times10^{-13}$\\
$F_\mathrm{2-10\,keV}$ (cgs) & $\left(1.48\pm0.07\right)\times10^{-12}$\\
$L_\mathrm{0.5-2\,keV}$ (cgs) & $\left(1.3\pm0.1\right)\times10^{41}$\\
$L_\mathrm{2-10\,keV}$ (cgs) & $\left(2.9\pm0.1\right)\times10^{41}$\\
\hline
\end{tabular}
\end{center}
\end{table}

We have also assessed whether the presence of a warm absorber is required by the
data. Unfortunately, the RGS spectra have a signal-to-noise ratio
too poor to perform any meaningful analysis. We
therefore tried to add a simple warm absorber (model \textsc{absori}\footnote{We are aware that this model oversimplifies the physics of warm absorbers, but here it is used only to assess the presence of absorption by ionised material.}
in \textsc{Xspec}) to the best fit model of the EPIC spectra. The
addition of this component is not required by the data and the value of the
ionization parameter $\xi$ is unconstrained. On the other
hand, substituting the neutral absorber included in the best fit
model with a warm absorber, we find the same column density and
$\xi<0.01$, confirming that the material is neutral.

\subsection{\label{opticaldata}Optical data}

\begin{figure*}
\epsfig{figure=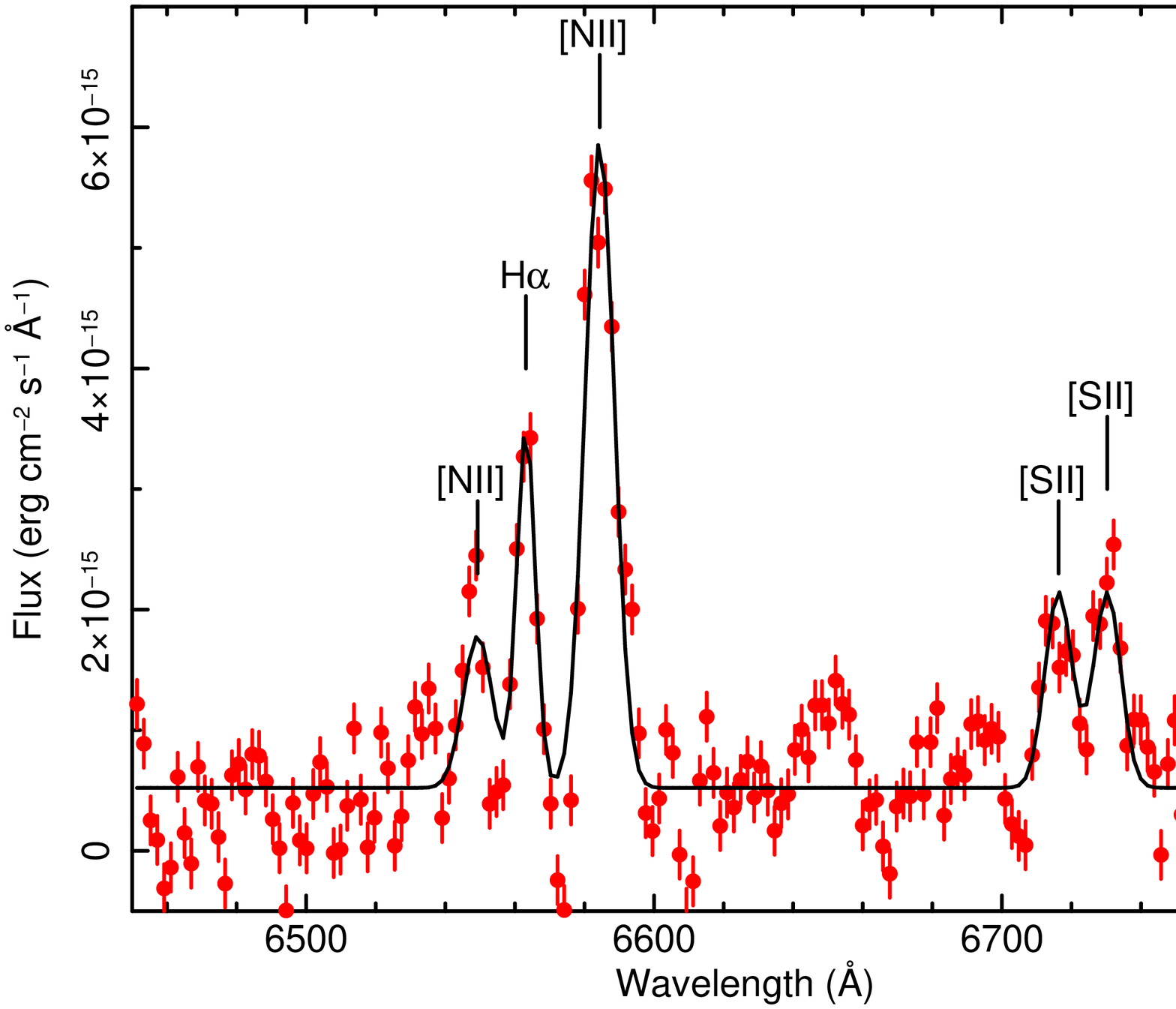,height=5.5cm}
\hspace{2cm}
\epsfig{figure=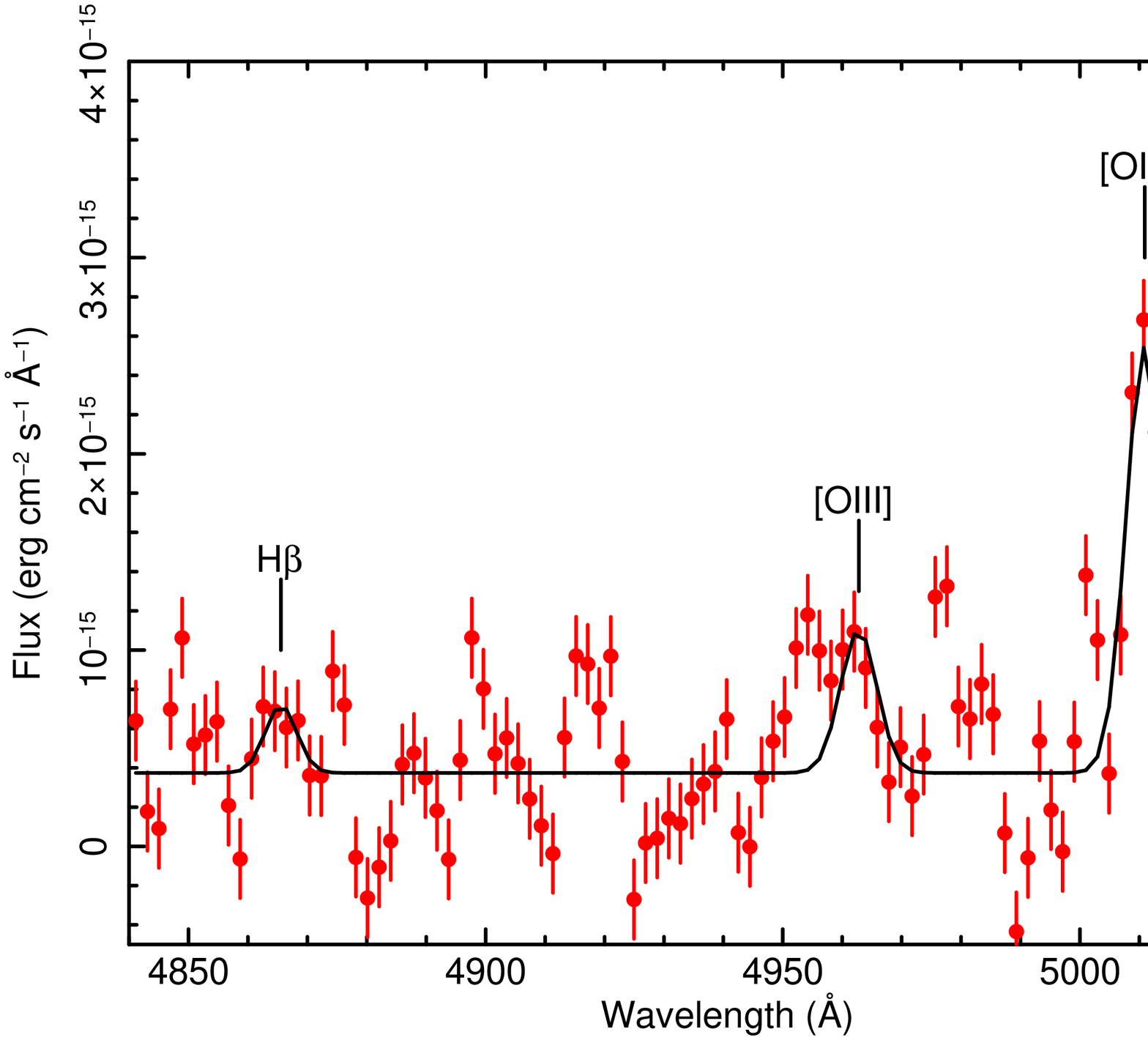,height=5.7cm}
\caption{\label{optical}NGC~3147 optical spectrum: $H_{\alpha}$ and $H_{\beta}$ spectral regions along with the best fit model.}
\end{figure*}

We measured the line
intensities and widths by fitting Gaussian profiles. The results are
presented in Table~\ref{table1}, the fitted spectra are in
Fig.~\ref{optical}. Line intensities are consistent with what found by
\citet{hfs97}, within errors. The FWHMs have been calculated taking into account the spectrograph
spectral resolution. Broad emission line components in the Balmer
lines are not required by the data, confirming previous
classification of NGC 3147 as a pure Seyfert 2 galaxy. In particular, the upper limit on the broad component of H$\alpha$ implies that its fractional contribution to the total (broad+narrow) H$\alpha$ emission would be at \textit{most} 5\%, while the average value in the \citet{ho97} sample is around 60\%. Note that the latter is a very conservative estimate, because the sample only includes low-luminosity Seyfert and LINER nuclei, mainly of intermediate type, where the broad component is expected to be much fainter than in bright Seyfert 1s. The lack of broad optical lines in NGC~3147 is therefore an intrinsic property of the source and not an artefact of low signal-to-noise.

We find a value for the Balmer decrement
of H$\alpha$/H$\beta=8\pm3$, consistent with the one ($\sim$ 5) measured by \citet{hfs97}. Adopting the latter, the [O III]$\lambda$5007 corrected flux is $\sim85\times\,10^{-15}{\rm \,erg\,cm^{-2}\,s^{-1}}$. Reddening insensitive
line ratios as measured from our spectrum are also consistent with
those from \citet{hfs97}: we measure [O III]$\lambda$5007/H$\beta=
7\pm2$ and [N II]$\lambda$6583/H$\alpha2.8\pm0.3$, while they
measure 6.1 and 2.7, respectively.

\begin{table}
\caption{\label{table1}Measured line parameters for the optical spectrum of NGC~3147. Line widths for [O\textsc{iii}], [N\textsc{ii}] and [S\textsc{ii}] doublets were fitted by tying down the velocity widths to the same value for each doublet. Values with $^*$ were kept fixed.}
\begin{center}
\begin{tabular}{lcc}
\hline
Line & Flux ($10^{-15}$ cgs) & FWHM (km s$^{-1}$)\\
\hline
H$_\beta$ (narrow) & $2.4_{-0.7}^{+0.5}$ &  $280_{-10}^{+20}$\\
H$_\beta$ (broad) & $<3.8$ & $2000^*$\\
$\mathrm{[{O\,\textsc{iii}}]}\,\lambda4959$ & $5.6_{-0.5}^{+0.7}$ & $370_{-40}^{+100}$\\
$\mathrm{[{O\,\textsc{iii}}]}\,\lambda5007$ & $17\pm2$ & $370_{-40}^{+100}$\\
H$_\alpha$ (narrow) & $20\pm2$  & $230_{-30}^{+10}$\\
H$_\alpha$ (broad) & $<1.1$  & $2000^*$\\
$\mathrm{[{N\,\textsc{ii}}]}\,\lambda6548$ & $12\pm2$ & $410\pm20$\\
$\mathrm{[{N\,\textsc{ii}}]}\,\lambda6583$ & $56\pm2$ & $410\pm20$\\
$\mathrm{[{S\,\textsc{ii}}]}\,\lambda6716$ & $16\pm2$ & $370_{-20}^{+40}$\\
$\mathrm{[{S\,\textsc{ii}}]}\,\lambda6731$ & $16\pm2$ & $370_{-20}^{+40}$\\
\hline
\end{tabular}
\end{center} 
\end{table}

\section{Discussion}

The best fit model for the XMM-\textit{Newton} spectra is that of a
typical type 1 AGN. The amount of neutral column density measured in
excess of the Galactic one is of the same order of the latter, and
it is therefore fully consistent with being due to the host galaxy's interstellar medium.
The powerlaw index, although somewhat flatter than the average type
1 radio-quiet object, is well within the large dispersion of the
hard X-ray $\Gamma$ distribution \citep[e.g.][]{pico05}. The EW of
the neutral iron K$\alpha$ line is fully consistent with those
typically found in unobscured AGN, especially given its low
luminosity \citep[e.g.][]{bianchi07}. The only peculiar aspect of
its X-ray spectrum is represented by the Fe
\textsc{xxv} emission line, whose EW is larger than the one typically arising in
a Compton-thin, photoionised gas \citep[see e.g.][]{bm02,bianchi05}. However, the line is not
statistically very significant and the errors on the EW are quite
large (see Sect. \ref{xmmanalysis}).

To test further the consistency of NGC 3147 being a type 1 AGN, we used
the QSO template from \citet{elvis94} and scaled it to the X-ray
flux measured by our X-ray observation. The predicted continuum
level at optical wavelengths is fully consistent with the measured nuclear spectrum. In
addition, we have examined the Optical Monitor \citep{Mason01} data
of our XMM-\textit{Newton} observation, where the nucleus of
NGC~3147 is detected as a point source. Once deconvolved the disk contribution and corrected for Galactic reddening, we get 
fluxes of $(1.81\pm 0.14)$ and $(0.94\pm0.14)\times 10^{-15}$ erg cm$^{-2}$ s$^{-1}$ \AA, at 2910 and 2310 \AA, respectively, leading to a two-point spectral index $\alpha_{ox}\simeq-1.33$\footnote{We adopt the original definition by \citet{tan79} as the ratio between X-ray and UV flux densities: $\alpha_{ox}=-0.384 \log[f(2\,\mathrm{keV})/f(2500\,\mathrm{\AA})]$.}, as found for most Seyfert 1s \citep[e.g.][]{stra05}.

However, the analysis of the \textit{simultaneous} optical spectrum presented in this paper
confirms the lack of broad permitted lines in this source. The very low column density measured in the X-rays
($N_H<5\times10^{20}$ cm$^{-2}$ at the 99\% confidence level for two
interesting parameters, corresponding to $\mathrm{A_V}<0.3$) is at odds with the large amount of
dust required to obscure the BLR. Generally, it is found that any deviation from the Galactic gas-to-dust ratio in AGN always goes in the opposite direction: the obscuration by dust is \textit{lower} than what expected from the associated gas column density \citep{mai01}. It is difficult to find a physical mechanism able to suppress gas without destroying dust.

If the source were Compton-thick, the lack of X-ray absorption could
be apparent, because we would not observe the primary emission at
all, but only the reflected light, unaffected by any line-of-sight
absorption. This solution is untenable for NGC~3147, because of the EW of the neutral iron line, much smaller than the one expected in the case of a reflection-dominated object \citep[$\simeq1$ keV: see e.g.][]{mbf96}. However, the (relatively) large EW of the Fe\textsc{xxv} line may be a signature of reflection from ionised material. In this case, the reprocessed spectrum would be indistinguishable from the primary powerlaw, but an Fe\textsc{xxv} EW as low as some hundreds of eV would be likely accompanied by detectable emission from Fe\textsc{xxvi} or lower ionised species \citep[see e.g.][]{bm02}. This solution would require a rather ad hoc geometry: a Compton-thick neutral absorber blocks the primary emission, while the reprocessed spectrum is instead dominated by an highly ionised material. Moreover, the high ratio between the 2-10 keV and the [O\textsc{iii}] reddening-corrected flux ($\simeq20$) strongly rejects the hypothesis that we are not directly observing the primary emission \citep[in this case, a ratio lower than 1 is expected: see e.g.][and references therein]{pb02}. Therefore, the source is quite unlikely to be Compton-thick. 

The only solution left is that the source intrinsically lacks the BLR. In this respect, it is interesting to note that NGC~3147 has no hidden BLR (HBLR) in polarised light (Tran, private communication). It was shown that HBLR sources are characterized, on average, by larger luminosities than non-HBLR ones \citep[e.g.][]{gh02}. Recently, \citet{es06} presented a model which depicts the torus as the inner region of a clumpy wind outflowing from the accretion disc. A key prediction of this scenario is the disappearance of the BLR at very low bolometric luminosities. Adopting a constant bolometric correction of 20 \citep[][but any other choice is irrelevant for the present estimate]{elvis94} to the observed X-ray luminosity, the bolometric luminosity of NGC~3147 is about $5\times10^{42}$ erg s$^{-1}$, which is somewhat larger than the `threshold' calculated by \citet{es06} and, in any case, larger than other low-luminosity objects showing broad lines \citep{ho97}. Moreover, the disappearance of the BLR \textit{follows} that of the torus, at lower luminosities. Therefore, their model predicts the absence of the torus in NGC~3147, requiring the strong neutral iron line to be produced in the accretion disc. Current data cannot confirm nor reject this hypothesis.

Another possibility is that the BLR is intimately linked to the accretion rate, rather than the luminosity, of the AGN, being formed in accretion disk instabilities occurring around a critical radius \citep{nic00}. Below a minimum accretion rate, this radius falls below the innermost stable orbit and the BLR cannot form. We found in the recent literature two estimates for the black hole mass of NGC~3147: $6.2\times10^{8}$ M$_{\odot}$ \citep[][inferred from the mass-velocity dispersion correlation]{merl03} and $2.0\times10^{8}$
M$_{\odot}$ \citep[][using a mass-$K_s$ bulge luminosity relation]{dd06}. Combined with the above-derived bolometric luminosity, we get a very low Eddington rate for NGC~3147, roughly ranging between $8\times10^{-5}$ and $2\times10^{-4}$. In any case, this is well
below the threshold (around $1\times10^{-3}$) proposed by \citet{nic00} and \citet{nic03}, thus supporting this scenario. 

In conclusion, it seems inescapable that the key parameter for the observational properties of this source is not orientation, but an intrinsic feature (be it low accretion rate or luminosity), which prevents the BLR to form. NGC~3147 is therefore a ``true'' Seyfert 2 without the BLR, the first unambiguous component of a new class of AGN which requires to be fitted in a more general Unified Model.

\section*{Acknowledgements}

This investigation benefited from an Italian-Spanish Integrated
Action, reference HI2006-0079. SB and GM acknowledge financial
support from ASI (grant 1/023/05/0). AC, FP, XB and FJC
acknowledge financial support by the Spanish Ministerio de
Educaci\'on y Ciencia under project ESP2006-13608-C02-01. We thank the referee, Luis C. Ho, for constructive suggestions, and Hien Tran and Andy Fabian for useful discussions.

\bibliographystyle{mn}
\bibliography{sbs}

\label{lastpage}

\end{document}